\documentclass[review]{elsarticle}

\usepackage{hyperref}
\usepackage[utf8]{inputenc} 
\usepackage[T1]{fontenc}    
\usepackage{hyperref}       
\usepackage{url}            
\usepackage{booktabs}       
\usepackage{amsfonts}       
\usepackage{nicefrac}       
\usepackage{microtype}      
\usepackage{amsfonts,amsmath,amsthm,array,amssymb}

\usepackage{graphicx}
\usepackage{setspace}
\usepackage{appendix}
\usepackage{xcolor}
\usepackage{lineno,hyperref}
\usepackage{natbib}
\journal{LETTERS IN BIOMATHEMATICS}
\nolinenumbers
\date{}
\begin{document}

\begin{frontmatter}

\title{Balancing Fiscal and Mortality Impact of SARS-CoV-2 Mitigation Measurements}


\author[mymainaddress,mythirdaddress]{Mayte\'e Cruz-Aponte \corref{mycorrespondingauthor}}
\cortext[mycorrespondingauthor]{Corresponding author}
\ead{maytee.cruz@upr.edu}

\author[mysecondaryaddress,forth,mythirdaddress]{Jos\'e Caraballo-Cueto}
\ead{jose.caraballo8@upr.edu}

\address[mymainaddress]{Department of Mathematics-Physics}
\address[forth]{Department of Business Administration }
\address[mysecondaryaddress]{Director, Census Information Center}
\address[mythirdaddress]{University of Puerto Rico at Cayey, Cayey, PR 00737}

\begin{abstract}
An epidemic carries human and fiscal costs. In the case of imported pandemics, the first-best solution is to restrict national borders to identify and isolate infected individuals. However, when that opportunity is not fully seized and there is no preventative intervention available, second-best options must be chosen. In this article we develop a system of differential equations that simulate both the fiscal and human costs associated with different mitigation measurements. After simulating several scenarios, we conclude that herd immunity (or unleashing the pandemic) is the worst policy in terms of both human and fiscal cost. We found that the second-best policy would be a strict policy (e.g. physical distancing with massive testing) established under the first 20 days after the pandemic, that lowers the probability of infection by 80\%. In the case of the US, this strict policy would save more than 239 thousands lives and almost \$170.8 billion to taxpayers when compared to the herd immunity case. 

\end{abstract}


\end{frontmatter}

\nolinenumbers

\section{Introduction}
 During the COVID-19 pandemic, many policymakers are usually facing two separated sources of information: economic models that usually predict an economic collapse \cite{JPMorgan} and epidemic models that focus on death counts \cite{Nishiura}. However, both the economic and mortality figures are key policy variables during a pandemic but few articles integrate both approaches \cite{eichenbaum, Alon}. In particular, no research (to our knowledge) has analyzed both the fiscal and mortality impact of different mitigation measurements. In this article we strive to fill that gap by approximating the impact of physical distancing and patient care on the death toll and government budget, in a attempt to find the optimal conditions to balance it all.
 
 Vaccination or therapeutics can eradicate epidemics from the population, like the case of smallpox \cite{bazin2000eradication,behbehani1983smallpox} but when a newly discovered virus hits the population, the entire world is at risk because everyone is susceptible as in the case of the novel SARS-CoV-2 that is impacting us since 2020 \cite{lai2020severe}. In the case of an imported infection (i.e. not an endemic epidemic), the first-best strategy would be to control borders, identify, treat and isolate infected individuals. This occurred in the U.S. with the Ebola virus, which never became an epidemic \cite{EbolaFlawless}. But when a virus is already circulating in a territory and there is no antidote or massive testing and contact tracing available, social or physical distancing is an alternative to mitigate a pandemic and provide the scientific community time to research and find alternative measures such as an effective treatment or a vaccine. Also, physical distancing measurements give fragile healthcare systems the leverage to take care of chronically ill patients without saturation of existing capacity. What are the fiscal and human costs of all these measurements in the short and long run?
 
 Thus, two research questions drive this study: What is the optimal physical distancing policies in a country and what are the implications of these policies for both the government budget and loss-of-life? We constructed an enhanced mathematical SIR (Susceptible, Infected, Recovered) epidemic model \cite{brauer2008compartmental} to simulate the COVID-19 epidemic in the US in an attempt to estimate the fiscal impact and the optimal conditions to mitigate this ongoing pandemic. We found that a policy of no physical distancing or a race towards herd immunity is not the optimal policy choice when both human and fiscal costs are considered. 
 
 In Section 2 we lay out our methodology. In Section 3 we show the dynamics associated to our calibrated system of differential equations. In Section 4 we discuss our results and in Section 5 we conclude and recommend public policies.

\section{Methodology}
We first describe a simple economy with three economic sectors; businesses, government, and a household sector that has two actors. In the second part of this section we describe our epidemic model.

\subsection{A Simple Economy}
In this economy, the household sector is mobile within the country (i.e. internal migration is allowed) and is composed of both L workers and U individuals that are not working. Thus, employment is not at its maximum level (in the economics jargon: employment is less than full). We follow \cite{Caraballo-Cueto2017} where the technology level (i.e. sophistication level of businesses) in period t–1 represents a barrier to entry in period t and prevent the economy to achieve full employment. This characterization allows us to consider the supply side shocks (i.e. disruption in the production process associated to inputs such as labor) associated to the COVID-19 pandemic \cite{guerrieri2020macroeconomic}, where laborers are impeded to work at the pre-pandemic level because of lock-downs or infections affecting members of the household sector. 

For simplicity, firms or businesses produce a number of goods, which require the only factor of production (i.e. labor) L. However, firms are able to adjust its output when external changes hit the labor supply (e.g. a pandemic that causes workers to reduce their working hours). The total output that considers the impact of such external changes is observed in, 
$Y_t = y L_t(1+H_t)$ 
where y is the technology level and H represents the external shock to the labor supply in period t.

We hold the following assumptions over H:
\begin{itemize}
    \item if physical distancing is implemented at t=1, $H_t=-0.3$ during the physical distancing. When the physical distancing ends in t=n, $H_{t=n} = 0.1$ this setting allows us to capture the V-shape growth that is being projected \cite{IMF} in the post-COVID-19 period.
    \item if no physical distancing is implemented, the pandemic ends in t=n+j, $H_{t=n+j+1} = 0$, and $H_{t<j} = -0.1$. This decline is lower than the lower bound estimate of fourteen percent decline in output constructed by JP Morgan for the U.S., a country that did not declare a general lock-down \cite{JPMorgan}. In our case, n+j+1 equals 500 days.
    \item if physical distancing is implemented late at t=i, with j<i,{ $H_t=-0.3$ during the physical distancing, $H_{t=n} = 0.1$ when the physical distancing gets relaxed, the pandemic ends in t=n+i. In our case, n+i equals 600 such that $H_{t>600} = 0$}. 
\end{itemize}

On the other hand, the government sector has a fixed level of expenditures predetermined in period t-1, that includes money transfers to low-income U. Tax revenues C are governed by,   

$C$ = $\tau Y_t$

where $\tau$ is the tax rate.
If there is a pandemic, the government spends $M_v$ in the treatment of each infected individuals.

\subsection{Epidemic model}

In order to assess the fiscal impact that a pandemic such as COVID-19 will convey by the epidemic dynamics we construct an epidemic model, with the addition of a fiscal impact differential equation based on the previous section that quantifies the impact in the budget of the country.  The epidemic model is a typical SIR-type model with the epidemiological classes needed for the evolution of COVID-19 within the population. There are six epidemiological classes defined Susceptible ({\textbf{S}}) individuals that progress to incubate the virus in the Exposed ({\textbf{E}}) class when they have an effective contact with an infected symptomatic individual ({\textbf{I}}) with probability $\frac{\beta SI}{N}$ using mass action or if a Susceptible individual had contact with an Asymptomatic ({\textbf{A}}) individual with a lower probability of infection than a contact with a symptomatic governed by the factor $\frac{\beta S \mu A}{N}$.  Treated individuals ({\textbf{T}}) are assumed to be quarantined, hence, they are not considered to be able to transmit the diseases in the context of this modelling approach. 
After 14 days of incubating the virus, a proportion of individuals becomes symptomatic and the rest are asymptomatic (estimated here at 35\% \cite{CNN35}).  Infected individuals that are asymptomatic recover without further complications, symptomatic individuals that get critically ill seek treatment while the mild cases recover. Following the related literature cited in Table \ref{Table:Parameters} below, symptomatic and treated individuals have a probability of dying due to health complications estimated at 3\%. In Table \ref{Table:Parameters} below there is a description of the parameters used. 

\begin{table}[h!]
\begin{tabular}{|c|p{4cm}|p{3.7cm}|p{1.8cm}|}
\hline 

\bf{Parameter}  & \textbf{Description}                                            & \textbf{Value} & \textbf{Reference} \\ \hline 
$\beta$    & Rate of infection                                      &      0.8  &      \cite{prem2020effect}     \\ \hline
$f(t)$    & Reduction of infection due to measures implemented over a time period t.                                      &      (0,1] refer to Equation \ref{Eqn:picewise}  &      Variable     \\ \hline
$\mu$      & Increase or decrease of asymptomatic rate of infection &      0.5 &     \cite{eikenberry2020mask}       \\ \hline
$\alpha$   & Incubation period                                      &  14 days     &       \cite{prem2020effect}         \\ \hline
q          & Probability of been asymptomatic                        &      35\% & \cite{CNN35}        \\ \hline
$\gamma$   & Recovery rate of asymptomatic                          &  14 days      &        \cite{prem2020effect}        \\ \hline
$\epsilon$ & Rate of hospitalization (critical illness)             &      20\% &      \cite{eikenberry2020mask}        \\ \hline
$\phi$     & Recovery rate after treated (hospital discharge)       &      14 days &    \cite{eikenberry2020mask}          \\ \hline
$\sigma$   & Recovery rate (mild cases no hospitalization needed)   &     14 days  &     \cite{eikenberry2020mask}         \\ \hline
$\delta$   & Death rate due to illness                              &     3\%/365 &    \cite{prem2020effect}       \\ \hline
$\tau$   & Tax rate                              &  0.24     &  \cite{0bbc27daen}          \\ \hline
$Y$   & GDP                              &  \$21.73 trillion divided by 365 days     &  BEA        \\ \hline
$H_t$   & Changes in the labor stock                              &  0.5 during physical distancing, 1.06 afterwards or 
0.7 if no physical distancing is ever implemented)     &    Assumptions described in section 2.1      \\ \hline
$M_v$   & Cost of treatment                              &  \$950 daily &  \cite{KFFHospitalized}
 \\ \hline
$M_p$   & Money transferred by the government to low-income individuals                         &   \$61 daily    &   \cite{cboFedSpending}      \\ \hline
$P$   & Fraction of dead individuals who received money transfers from the government                &    80\% of victims received transfers. & \cite{thompson2011risk,tricco2012impact}
\\ \hline
\end{tabular}
\caption{Parameters of the Epidemiological and Fiscal dynamic model for COVID-19 and the USA budget.}
\label{Table:Parameters}
\end{table}

\subsection{System of Ordinary Differential Equations}

{The system of differential equations for COVID-19, is now presented in Equations \ref{system:S} to \ref{system:R}. In particular, the system is described as:

\begin{eqnarray} 
\dot{S} &=& -\frac{\beta S (I + \mu A)}{N} \label{system:S} \\
\dot{E} &=&  \frac{\beta S (I + \mu A)}{N} - \alpha E \label{system:E} \\
\dot{A} &=&  \alpha q E - \gamma A \label{system:A} \\
\dot{I} &=&  \alpha (1-q) E - (\epsilon + \delta+ \sigma)I \label{system:I} \\
\dot{T} &=&  \epsilon I - (\delta + \phi) T \label{system:T} \\
\dot{R} &=&  \sigma I + \phi T + \gamma A \label{system:R}\\
\dot{D} &=& \delta (I+T) \label{system:death}\\
\dot{G} &=& \tau y H_t - M_v  T + M_p P*D \label{impactG}
\end{eqnarray}}

Equation \ref{impactG} states the changes in the government budget $G$ caused by the tax revenues $C$ collected in time $t$, minus the cost of treatment $M_v$ multiplied by the number of treated individuals $T$ in time t, plus the cost of money transfers $M_p$ that the government invested to help the fraction P of dead individuals $\delta (I+T)$ that had low incomes. In other words, the last terms accounts for the fiscal savings that the government obtain when poor individuals die during the pandemic.

The parameter values in our table were taken from a recollection of the events developing on COVID-19 and the literature, as cited in Table 1. The rate of infection (0.8), the incubation period (14 days), the recovery rate of asymptomatic (14 days), and the death rate due to illness (3\% divided by 365) were obtained in \cite{prem2020effect}. The probability of being asymptomatic (35\%) comes from CDC estimates \cite{CNN35}. The increase or decrease of asymptomatic rate of infection (0.5), the rate of hospitalization (20\%), and the recovery rates of both mild cases where no hospitalization is needed and after being discharged from hospital (14 days) were taken from \cite{eikenberry2020mask}. The reduction of infection due to measures implemented over a time period $t$ varies from (0,1] where 1 means there are no measures implemented and a value less than one represent lower infection rates due to the measures implemented over time as defined by equation \ref{Eqn:picewise}. The tax rate was obtained by dividing the 2019 tax revenues by the 2019 Gross Domestic Product (GDP). The daily GDP, Y, was obtained by dividing the US 2019 by 365 days. The cost of treatment was the lowest found in \cite{KFFHospitalized}, where less-severe hospitalization cost approximately \$13,300, divided by 14 days that are the average days on hospitalization) . The money transferred by the government to low-income individuals was taken from the federal average yearly payment to 65 years or more individuals in \cite{cboFedSpending}, which is approximately \$22,265 divided by 365 days. The fraction of dead individuals who received money transfers from the government was assumed to be 80\%, which is closed to the percentage of dead individuals that are elder, and we assume that there is no disproportion of low-income persons in other ages \cite{thompson2011risk,tricco2012impact}

Perturbations around the figures showed in Table \ref{Table:Parameters} may affect the quantitative magnitude of our figures, but results are qualitatively similar after such perturbations. In other words, our findings for the second-best scenarios are robust to deviations of our parameters: such deviations will largely scale up or down the quantitative aspects of each scenario.

\subsection{Basic reproductive number}

{The parameters of the model were fitted to maintain a basic reproductive number relatively close to $3.8$. For comparison purposes lets note that the $\Re_0$ of seasonal influenza ranges approximately within 1.7 to 2.1 \cite{truscott2009quantifying}. For the 2009 influenza A-H1N1 the $\Re_0$ was estimated to be between 1.2 to 1.6.} 

In particular, the basic reproductive number $\Re_0$ is the number of secondary cases a single infectious individual generates during the period of infectivity on a completely susceptible population.  We assume that  the entire population is susceptible such that $S \approx N$ and that the epidemic has not started in t=0. The individuals that can potentially infect the population in our model are infected individuals that are either symptomatic or asymptomatic. Treated individuals are assumed to be quarantined. Following the related literature \cite{van2002reproduction}, we use the next  generation operator to compute the $\Re_0$.

Using the differential equations that contribute to the spread of the infection within the population (Equations \ref{system:E}, \ref{system:A} and \ref{system:I}). We define the vector $F$ to be the rate of new infections flowing to the latent compartment and the vector $V$ to be the rates of the equations that denote  the transfer of individuals within the compartment that are able to transmit the disease with opposite signs.  Hence, we define
\\
\begin{center}
$F = \left[ \begin {array}{c}  \beta \frac{S(I+\mu A)}{N}\\ 
\noalign{\medskip}0
\\ \noalign{\medskip}0\end {array} \right]$ 
and 
V = $\left[ \begin {array}{c}   
\alpha E \\  
\noalign{\medskip} -q \alpha E + \gamma A \\ 
\noalign{\medskip} -(1-q) \alpha E + (\epsilon + \delta + \sigma) I 
\end{array} \right]$.
\end{center}

In order to compute $\Re_0$, we compute the gradient of $F$ i.e. the partial derivatives of the vector defined as
$\mathfrak{F} = 
\left[ 
\frac{\partial F}{\partial E}, \frac{\partial F}{\partial A}, \frac{\partial F}{\partial I}
 \right]$
and the gradient of $V$ is computed as well defined as
$\mathfrak{V} = 
\left[ 
\frac{\partial V}{\partial E}, \frac{\partial V}{\partial A}, \frac{\partial V}{\partial I}
 \right]$
we get the following matrices: \\
\begin{center}
$\mathfrak{F} = 
\left[ \begin {array}{cccc} 0 & \mu \beta &\beta 
 \\ \noalign{\medskip}0&0&0\\ \noalign{\medskip}0&0&0\end {array} \right]$ 
and
$\mathfrak{V} = 
\left[ \begin {array}{cccc} \alpha &0&0\\ 
\noalign{\medskip} - q \alpha&\gamma&0\\ 
\noalign{\medskip} -(1 - q) \alpha&0& (\epsilon + \delta + \sigma)\end {array} \right]$ 
\end{center}

Now we compute the matrix $\mathfrak{FV}^{-1}$ in order to obtained the basic reproductive number $\Re_0$. Then, $\Re_0$ is then the spectral radius of the second generation operator matrix $\rho(\mathfrak{FV}^{-1})$.  By definition, the spectral radius of a square matrix is defined as the dominant eigenvalue or the largest eigenvalue when they are compared taking their absolute value. Hence, 
\\
\begin{eqnarray*} \mathfrak{FV}^{-1} &=&
\left[ \begin {array}{cccc} 0 & \mu \beta &\beta 
 \\ \noalign{\medskip}0&0&0\\ \noalign{\medskip}0&0&0\end {array} \right]
\left[ \begin {array}{cccc} \frac {1} {\alpha}&0&0
\\ \noalign{\medskip}{\frac {q}{\gamma}}&\frac {1} {\gamma}&0
\\ \noalign{\medskip}{\frac {1 - q }{ \delta + \sigma + \epsilon}} &0
& \frac {1} { \delta + \sigma + \epsilon} \end {array} \right] \\
 &=&
\left[ \begin {array}{cccc} 
\frac { \beta \left[ \mu q (\delta + \sigma + \epsilon ) - \gamma q + \gamma \right ] } {\gamma (\delta + \sigma + \epsilon )} &
\frac {\mu}{  \gamma}
&\frac {1}{(\delta + \sigma + \epsilon )}
\\ \noalign{\medskip}0&0&0
\\ \noalign{\medskip}0&0&0
\end {array} \right]
\end{eqnarray*}
and the dominant eigenvalue of $ \mathfrak{F}\mathfrak{V}^{-1}$ is $\rho(\mathfrak{FV}^{-1})=\frac { \beta \left[ \mu q (\delta + \sigma + \epsilon ) - \gamma q + \gamma \right ] } {\gamma (\delta + \sigma + \epsilon )}$, which means that the basic reproductive number is in fact:
 \begin{eqnarray}
\Re_{0}&=&\frac { \beta \left[ \mu q (\delta + \sigma + \epsilon ) - \gamma q + \gamma \right ] } {\gamma (\delta + \sigma + \epsilon )}
\label{ErreZero}
 \end{eqnarray}

Our intention on this article is to study the impact of the COVID-19 epidemic in an attempt to estimate the optimal conditions to mitigate the fiscal and mortality impact associated to this pandemic. Thus, the stability of the system and equilibrium points will not be addressed, only the $\Re_0$ was computed.  We focused our efforts on simulating scenarios, which are presented in the next set of sections.

\section{Physical distancing dynamics generalization}

{ In order to go through the methodology of our simulations we lay out simple scenarios where we represent the effect of implementing public health policies such as physical distance dynamics within our model. In order to simulate the effect of reducing the infection rate by lowering the contact rate within the population, we modulate the infection ratio $\beta$ with a time dependent piece-wise continuous function equation \ref{Eqn:picewise}, $f(t)$, that lowers the infection rate for time $t_{pdOn}$ where physical distancing policies are implemented and raise the infection rate after time $t_{pdOff}$. This late increase of the infection rate is not to its full force because we need to account for the measures taken by the population to prevent infections such as using mask and being more sanitized until a determined time $t^*$ that either restarts physical distance measures or is the end of the simulated period. See Appendix A for an example on the physical distancing mechanism.} 

\begin{eqnarray} 
f(t) =  \left\{
\begin{matrix}
      1, & 0\le t < t_{ pdOn }, & \text{No physical distancing }\\ 
[0.10,0.50], & t_{ pdOn }\le t \le t_{ pdOff }, & \text{ Physical distancing}\\ 
[0.75,0.90], & t_{ pdOff }<t<t^*,  & \text{Measures relaxed} \\
\end{matrix} 
\right.
\label{Eqn:picewise}
\end{eqnarray}
{

\section{Fiscal and Mortality implications under physical distancing scenario: The Case of the US}
We simulate the effects of varying infection rates in the total population of the US. In the case shown in Figure \ref{Fig:EbolaCase_LD_01_day2_50First_75Others_4On_8off}, we assume that distancing policies are implemented two days after the start of the epidemic. The grey line illustrates the effect of varying the probability of infection within the population: first the probability is reduced by 90\% for four weeks due to an extreme measure (e.g. because of a lock-down with quick massive testing), then is relaxed to 50\% for eight weeks, then two cycles of extreme measurement for four weeks followed by a relaxation of 25\% for 8 weeks and onward. This 25\% reduction in the original infection rate $\beta$ assumes that people are more careful and take personal decisions to avoid infections. The black line, on the other hand, represents the case when no physical distancing measures are ever implemented.

\begin{figure}[h]
\centering 
\includegraphics[width=\textwidth]{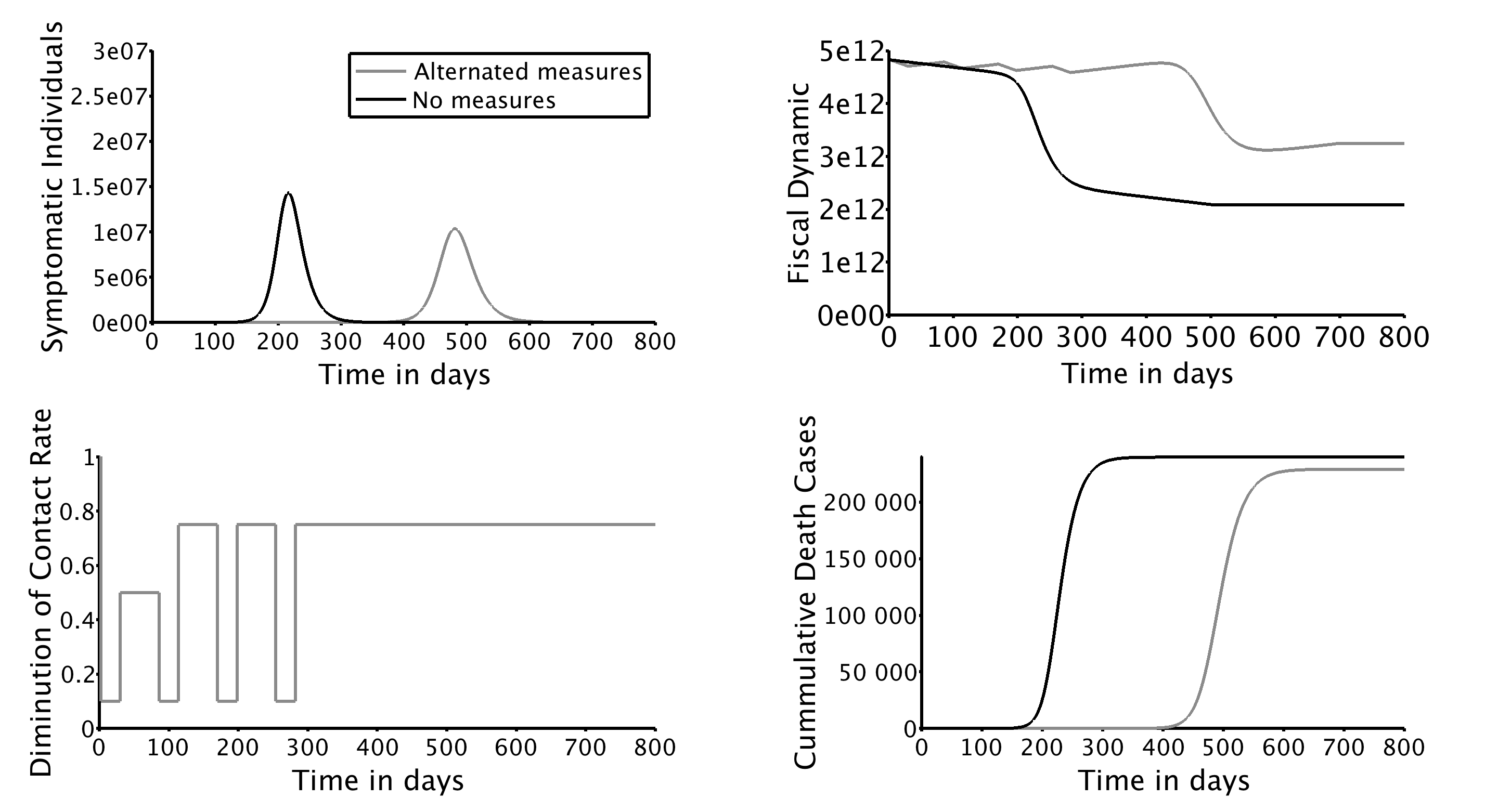}
\caption{Alternated physical distancing starting two days after the epidemic, as described on Table \ref{Table:EbolaCase_LD_02_day2_50First_75Others_4On_8off}.
 \label{Fig:EbolaCase_LD_01_day2_50First_75Others_4On_8off}}
\end{figure}

Fiscal and death figures are affected by the number of individuals that circulate in the economy.   The top left graph illustrates the symptomatic cases based on the modulation of the physical distancing measures, as shown in the bottom left graph. Note that in the case of no physical distance or herd immunity, the infections grow faster and earlier than in the modulated case, as shown by the black line. Because the government has to treat those cases in a fast-track basis, the fiscal impact of no physical distance is reflected earlier than in the case of alternated physical distance. Note that the economies obtained by the government when low-income individuals who receive money transfers die are not sufficient to offset the fiscal losses associated with the pandemic. Human costs also come up earlier in the case of no distancing, as shown in the bottom right graph where cumulative dead cases are illustrated in Figure  
\ref{Fig:EbolaCase_LD_01_day2_50First_75Others_4On_8off}.

In the long run, herd immunity has a higher death toll and implies more government expenditures than the alternated intensity of physical distancing, as shown in Table \ref{Table:EbolaCase_LD_02_day2_50First_75Others_4On_8off}. In particular, with no physical distancing 10,689 more individuals would die than in the alternated scenario and the government would lose \$1.16 trillion more than in the varying physical distancing. Note that under the alternated scenario scientists have approximately 400 days to find an antidote with a very low number of victims, vis-a-vis 150 days under the no distancing case. 
\begin{figure}[h]
\centering 
\includegraphics[width=\textwidth]{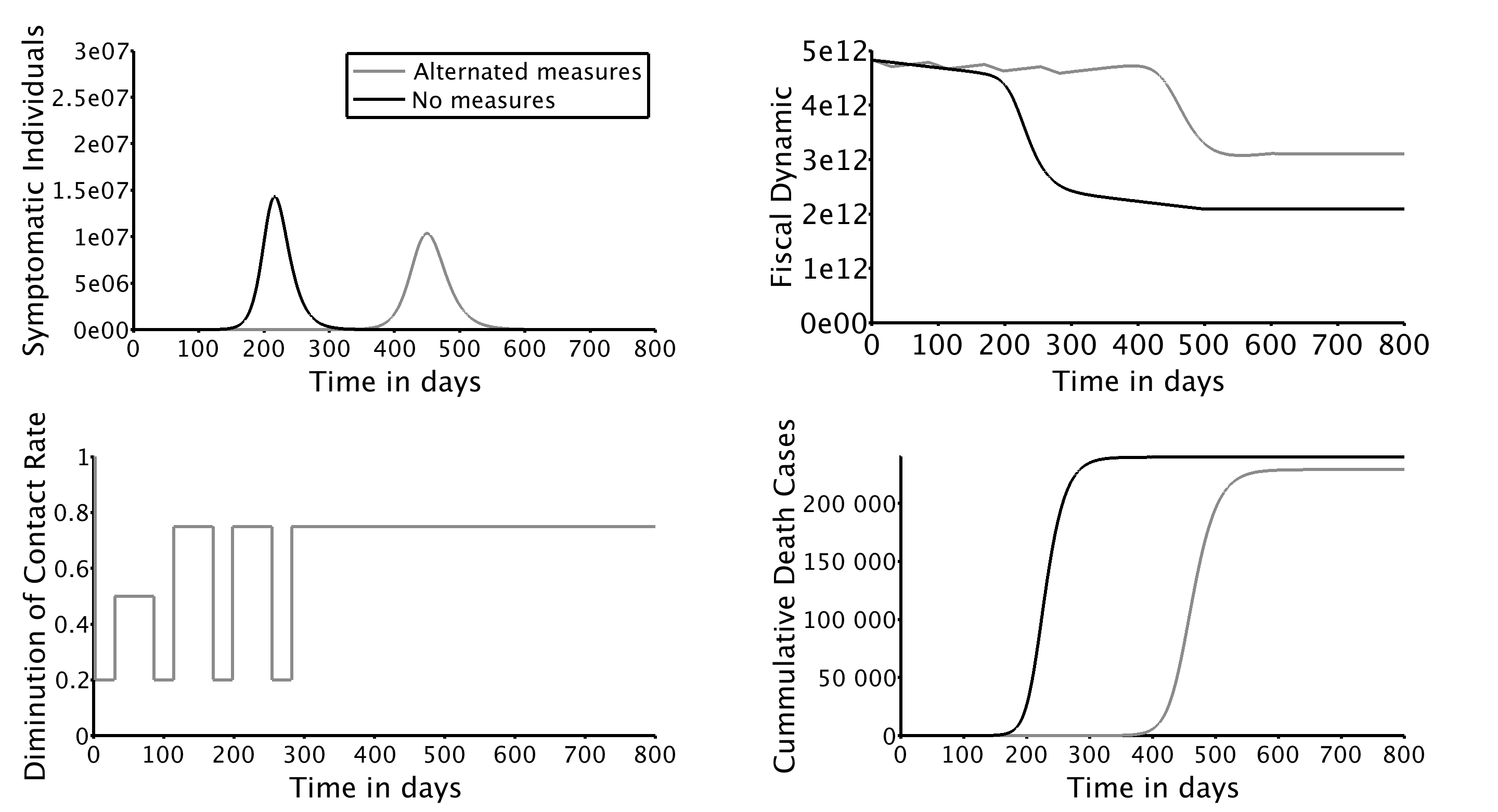}
\caption{Varying physical distancing starting two days after the epidemic: lowering the infection rate 20\% for four-week intervals and increasing it between 50\% to 75\% for eight weeks intervals as described on Table \ref{Table:EbolaCase_LD_02_day2_50First_75Others_4On_8off}.
 \label{Figure:EbolaCase_LD_02_day2_50First_75Others_4On_8off}}
\end{figure}
\begin{table}[h!]
\begin{center}
\begin{tabular}{|l|r|r|}
\hline
\begin{tabular}[c]{@{}l@{}}Reduction in probability of infection\end{tabular} & \begin{tabular}[c]{@{}l@{}}Cumulative \\ Death cases\end{tabular} & Budget after 800 days \\ \hline
No measures  & 239,646 	 & \$2,090,225,896,159   \\ \hline
20\% for 4 weeks,50\% for 8 weeks      
& 228,956 	 & \$3,106,451,634,585\\ 
20\% for 4 weeks,75\% for 8 weeks (twice) & & \\
75\% afterwards & &\\  \hline
10\% for 4 weeks,50\% for 8 weeks      
& 228,957 	 & \$3,249,326,371,355        \\ 
10\% for 4 weeks,75\% for 8 weeks (twice) & & \\
75\% afterwards & &\\    
\hline
20\% for 8 weeks,50\% for 4 weeks      
& 228,946 	 & \$2,609,322,982,259  \\ 
20\% for 8 weeks,75\% for 4 weeks (twice) & & \\
75\% afterwards & &\\  
\hline
60\% for 800 days &212,539 	 & \$464,708,870,047  \\ \hline
20\% for 800 days & 0 	 		 & \$2,260,976,930,754    \\ \hline

\end{tabular}
\caption{Cumulative death cases and fiscal impact of Figure \ref{Fig:EbolaCase_LD_01_day2_50First_75Others_4On_8off}, \ref{Figure:EbolaCase_LD_02_day2_50First_75Others_4On_8off}, and \ref{Fig:EbolaCase_LD_02_day2_50First_75Others_8On_4off}}
\label{Table:EbolaCase_LD_02_day2_50First_75Others_4On_8off}
\end{center}
\end{table}

{ On Figure \ref{Figure:EbolaCase_LD_02_day2_50First_75Others_4On_8off}}, we illustrate the simulated figures that are obtained when we shorten the length of the cycles showed previously. Here we are also holding the assumption that the pandemic started on day two. The variation in the probability of infection is shown in the gray line, which goes first to an extreme reduction  of 80\% for four weeks, then is relaxed to 50\% for eight weeks, then two cycles of extreme measurement for four weeks followed by a relaxation of 25\% for eight weeks and onward. The black line here also illustrates the case when no physical distancing measures are ever implemented. 

Similar to Figure \ref{Fig:EbolaCase_LD_01_day2_50First_75Others_4On_8off}, here the black line peaks first while the restriction on population mobility postpones and lowers the infection curve. The final amounts for cumulative deaths and the budget are presented on Table \ref{Table:EbolaCase_LD_02_day2_50First_75Others_4On_8off}. We observed that the strict alternated scenario of Figure \ref{Fig:EbolaCase_LD_01_day2_50First_75Others_4On_8off} would save the federal government \$142.9 billion more than in the relaxed scenario of Figure \ref{Figure:EbolaCase_LD_02_day2_50First_75Others_4On_8off}, while mortality is virtually the same. In other words, the distancing cycles shown in Table \ref{Table:EbolaCase_LD_02_day2_50First_75Others_4On_8off} appear to be less optimal than in the more restrictive case of Table \ref{Table:EbolaCase_LD_02_day2_50First_75Others_4On_8off}.

\begin{figure}[h]
\centering 
\includegraphics[width=\textwidth]{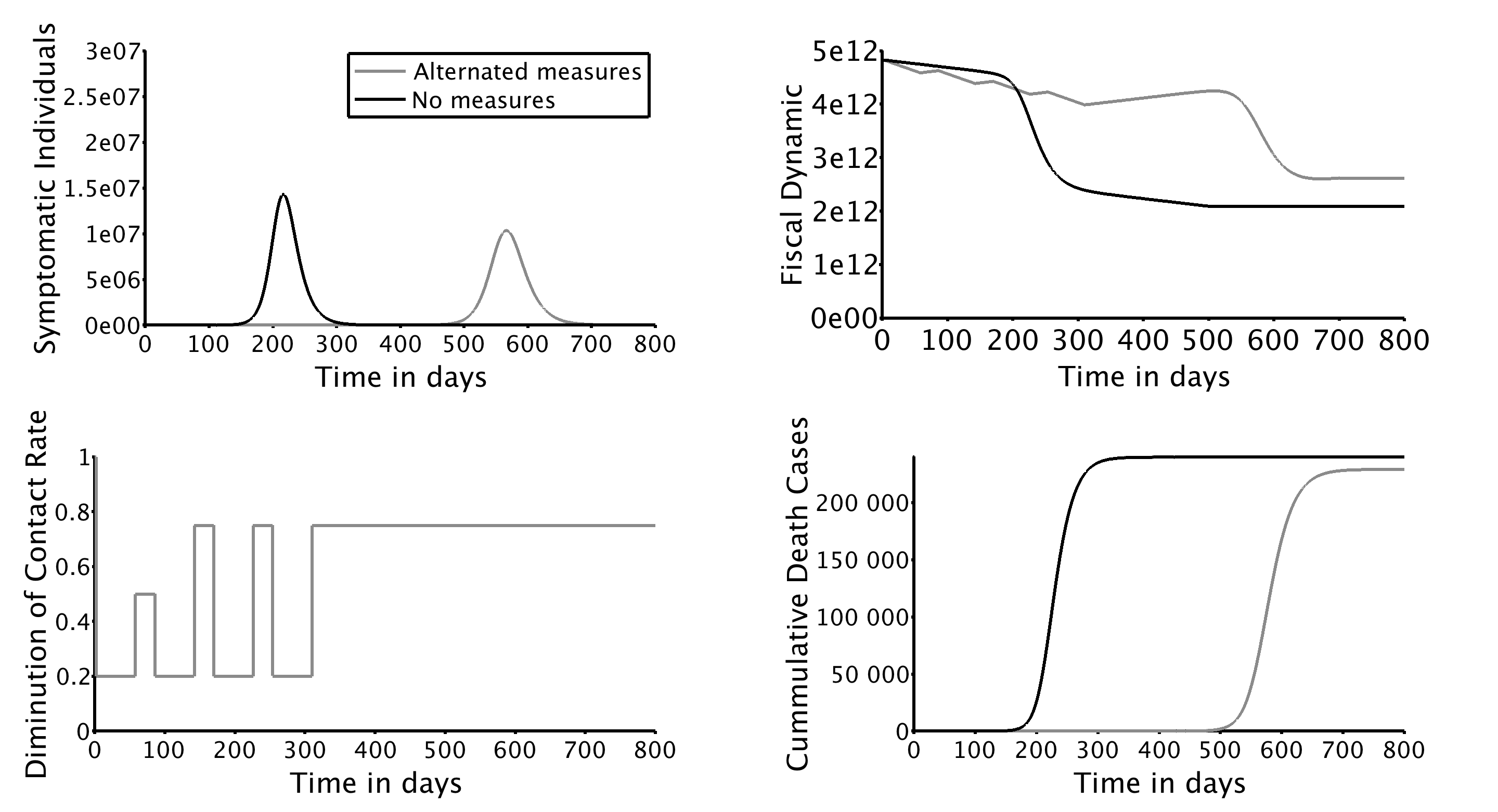}
\caption{Changing week length to the case of Figure \ref{Figure:EbolaCase_LD_02_day2_50First_75Others_4On_8off}}
 \label{Fig:EbolaCase_LD_02_day2_50First_75Others_8On_4off}
\end{figure}

What would be the effect of holding the same levels of distancing but changing the length under each regime? In particular, if we enhance the period under the restrictive infection and shorten the relaxation in each cycle, how would the death toll and the fiscal cost change? 

In Figure \ref{Fig:EbolaCase_LD_02_day2_50First_75Others_8On_4off} we observe that the peak of infection is postponed further when compared to the regime of Figure \ref{Figure:EbolaCase_LD_02_day2_50First_75Others_4On_8off}, leaving close to 100 days more for the development of an antidote. If no antidote is ever found in 500 days, under these enhanced cycles 10 fewer people would die than in the case of Figure \ref{Figure:EbolaCase_LD_02_day2_50First_75Others_4On_8off} where the same probabilities of infection are assumed. However, in the case of the cost that this pandemic represents to the federal budget, this restrictive regime of Figure \ref{Fig:EbolaCase_LD_02_day2_50First_75Others_8On_4off} costs \$497.1 more billion than in the cycles of Figure \ref{Figure:EbolaCase_LD_02_day2_50First_75Others_4On_8off}.

\begin{figure}[h]
\centering 
\includegraphics[width=\textwidth]{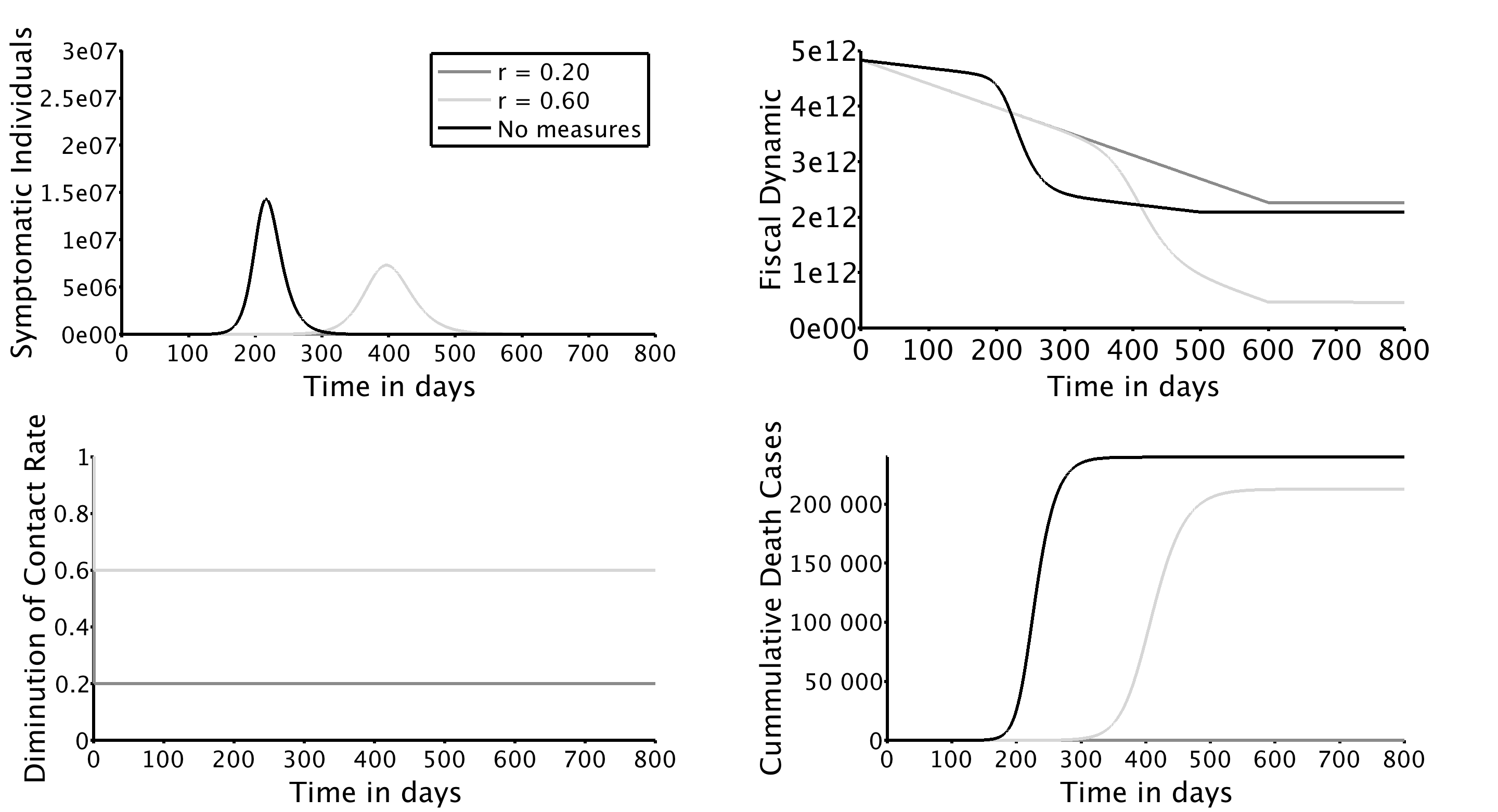} 
\caption{Effects of persistent distancing after day two of the pandemic}

\label{Fig:2040Day2}
\end{figure}
\begin{table}[h]
\begin{center}
\begin{tabular}{|c|r|r|}
\hline
\begin{tabular}[c]{@{}l@{}}Reduction in probability of infection\end{tabular} & \begin{tabular}[c]{@{}l@{}}Cumulative \\ Death cases\end{tabular} & Budget after 800 days \\ \hline
No measures 
 & 239,646 	 & \$2,090,225,896,159 \\ \hline
60\% for 800 days &212,539 	 & \$464,708,870,047  \\ \hline
20\% for 800 days & 0 	 		 & \$2,260,976,930,754    \\ \hline

\end{tabular}
\caption{Cumulative death cases and fiscal impact of persistent and early distancing of Figure \ref{Fig:2040Day2}}
\label{Table:2040Day2}
\end{center}
\end{table}

If we let the probability of infection to be lessened by 80\% of distancing to be in place for the whole period, then virtually no one would die out of COVID-19. This persistent measure would save 239,646 lives with respect to what would have occurred without health policies. However, if the same persistent distancing reduced the probability of infection by 40\%, mortality would be 27,107 lower than the herd immunity scenario.
In the case of the fiscal costs of the different alternatives,  budget is presented in Table \ref{Table:2040Day2}. Since there is no infection curve in the restrictive and permanent case, the budget of the government is mostly affected by the decline in tax revenues caused by lowered economic output. However, when infection rate is lowered just by 40\%, the government budget becomes highly impacted after day 400: once infection starts to increase, the public sector is affected by both spending more in treatment and by the reduction in tax revenues caused by the economic collapse. In the case of no physical distancing, even though economic output decreased, the federal budget is highly reduced by the treatment cost. At the end, if there is an opportunity to implement a persistent and early public health policy, it would be better to do it intensely (e.g. by combining a strong physical distance with massive testing): otherwise, the relatively human and fiscal cost would increase significantly.

How would the figures change if the persistent measure is implemented late? The evolution of infection and mortality is similar, but there are differences in the long run for the associated fiscal costs. When the infection is reduced by 40\% or 80\%, the associated cost of the pandemic to the government decreased by \$51.4 billion than when the public health policy is implemented earlier like in Figure \ref{Fig:2040Day2}. This difference arises because the economy was not halted during the first weeks of the pandemic. That is, even in the case of these late decisions the government is better off in implementing the strict measure: death toll does not change while fiscal costs perform much  either on day 2 or day 20 after the pandemic (See Appendix B).  

In Figure \ref{Fig:10_8weeks_50_24weeks} we show the effects of lowering the probability of infection within the whole U.S. population for different time intervals. When infection is reduced by 90\% for eight weeks, mortality is 10,713 lower than in the herd immunity case and the fiscal cost of the pandemic is the minimum with respect to any of the scenarios presented in this article. That is, this short and strong distancing policy would have a lower death toll (24 deaths) than in the alternated scenario of Figure \ref{Fig:EbolaCase_LD_01_day2_50First_75Others_4On_8off}, but with a lowered fiscal cost (\$180.3 billion less).
\begin{figure}[h]
\centering 
\includegraphics[width=\textwidth]{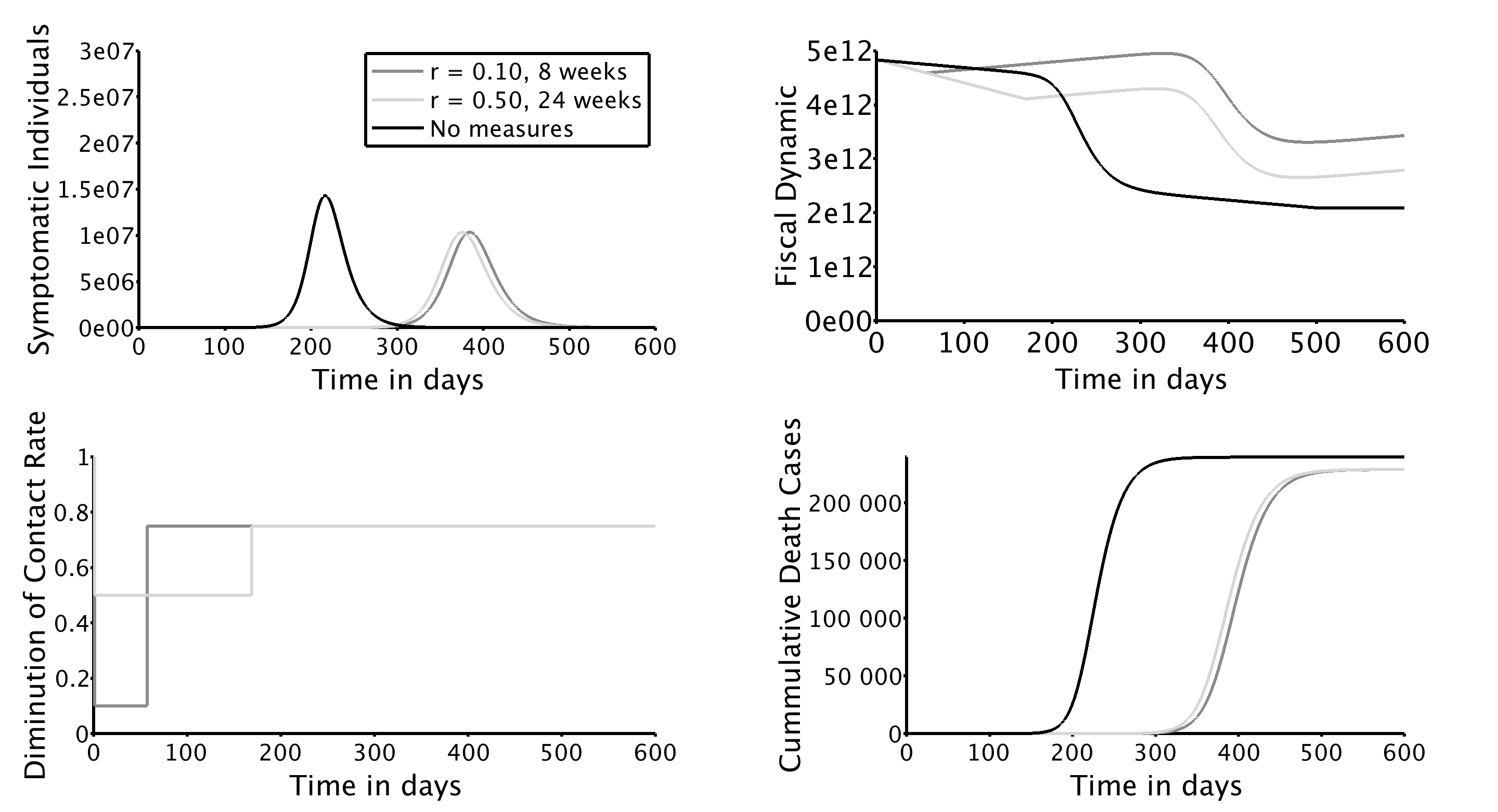}
\caption{Varying probability of infection between 10\% and 50\% with different time intervals, two days after the pandemic}
\label{Fig:10_8weeks_50_24weeks}
\end{figure}
\begin{table}[h]
\begin{center}
\begin{tabular}{|l|r|r|}
\hline
\begin{tabular}[c]{@{}l@{}}Reduction in probability of infection\end{tabular} & \begin{tabular}[c]{@{}l@{}}Cumulative \\ Death cases\end{tabular} & Budget after 600 days \\ \hline
No physical distancing	
  & 239,646 	 & \$2,090,225,900,896  \\ \hline
10\% for 8 weeks   
& 228,933 	 & \$3,429,605,751,038  \\ \hline
50\% for 24 weeks   
& 228,941 	 & \$2,789,411,480,863   \\ \hline
\end{tabular}
\caption{Cumulative death cases and fiscal impact of varying probability of infection between 10\% and 50\% with different time intervals of Figure \ref{Fig:10_8weeks_50_24weeks}}
\label{Table:10_8weeks_50_24weeks}
\end{center}
\end{table}

If the infection rate is reduced by 50\% for twenty four weeks, the death toll is a bit higher than in the case of a strong policy for eight continuous weeks. In either case, the pandemic would kill at least 10,705 fewer individuals than in the herd immunity case. In terms, of budget, the longer period under distancing induces a larger fiscal decline than in the case of the eight-week distancing. However, the herd immunity is again the most costly option in terms of both death and government budget.

These results reveal that if a policymaker is going to implement distancing measures, overall better results would be found with a strict measure for a short period than with a weak measure for a long period. In fact, in terms of human and fiscal costs, we can state that this strong policy for eight continuous weeks is the preferred option after the case of Figure \ref{Fig:2040Day20}.
\begin{figure}[h]
\centering 
\includegraphics[width=\textwidth]{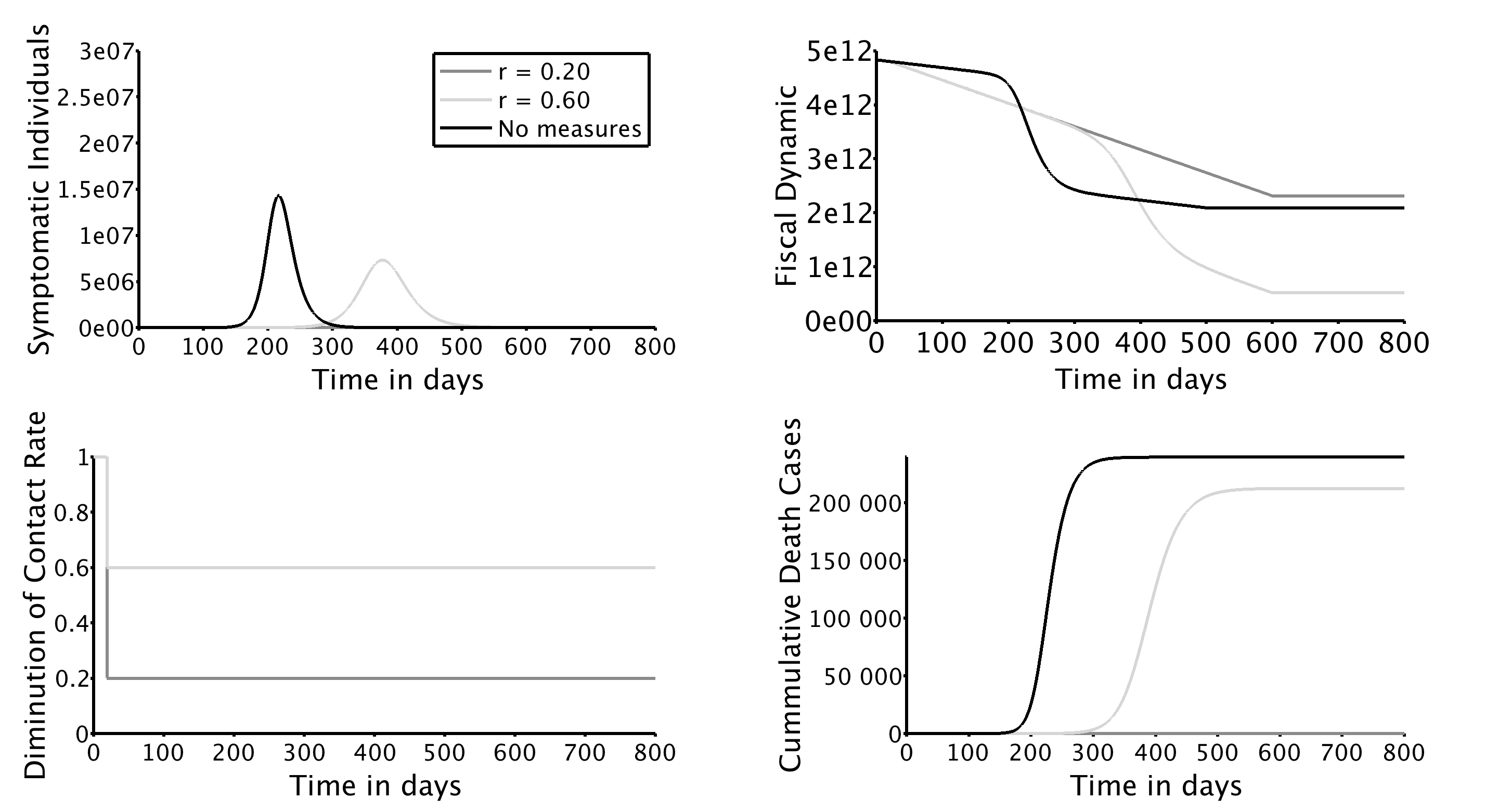} 
\caption{Effects of persistent distancing implemented in day 20 of the pandemic.}
\label{Fig:2040Day20}
\end{figure}
\begin{table}[h!]
\begin{center}
\begin{tabular}{|c|r|r|}
\hline
\begin{tabular}[c]{@{}l@{}}Reduction in probability of infection\end{tabular} & \begin{tabular}[c]{@{}l@{}}Cumulative \\ Death cases\end{tabular} & Budget after 800 days \\ \hline
No measures  & 239,646 	 & \$2,090,225,896,159 \\ \hline
60\%  & 212,539 	 & \$516,145,614,624    \\ \hline
20\%  & 0  	                  & \$2,312,415,608,494    \\ \hline

\end{tabular}
\caption{Cumulative death cases and fiscal impact of persistent but late distancing on Figure \ref{Fig:2040Day20}}
\label{Table:2040Day20}
\end{center}
\end{table}

\section{Conclusions and Policy Recommendations}
A challenge that policymakers face during a pandemic is to save lives at the minimum fiscal costs. In the case of COVID-19, the first-best policy to minimize human and fiscal costs would be reached by identifying, treating, and isolating incoming infected individuals. When this opportunity is missed, second-best policies need to be searched. We conclude that, when both fiscal and human costs are equally relevant, the second-best policy is reached when policies to significantly reduce the transmission rate are taken. In particular, if the transmission rate is lowered by 80\%, either on day 2 or day 20 after the beginning of the pandemic (see Appendix B), both human and fiscal costs associated to the pandemic are minimized with almost no dead cases and \$2.57 trillion in net impact to the government budget. These early policies can take the form of physical distancing combined with massive testing. 

The third best policy is found when a strong lock-down or similar policy reduced the infection rate by 90\% for eight weeks. If one focuses only on money disregarding life, this policy is the second-best policy: the pandemic would cost approximately \$1.4 trillion. The fourth best policy is to  alternate physical distancing measures: strict distancing for four weeks followed by a mild relaxation of eight weeks, then two cycles of returning to strict distancing for four weeks followed by eight weeks of relaxation. 

Unleashing the pandemic without taking any containment policy does not minimized the fiscal cost of the pandemic in any of our several simulations, except for the case when the transmission rate is reduced mildly. But even when compared to that case of a permanent reduction of 40\% to the infection rate, this herd immunity case still results in 27,107 more lives lost. This is because the race towards herd immunity always has the greatest human costs. In fact, at the time of our writing, scientists cannot clearly state that immunity is found after surviving the infection \cite{NPRImmunity,WhoImmunity}.

We do not consider what type of fiscal policy can be implemented to counteract the fiscal costs associated to the pandemic. Instead, we attempt to identify the cost of the pandemic and, from that amount, fiscal policies can be tailor-made to address specific needs.

We acknowledge that our model may not be exported to developing countries where the lock-down can also result in deaths of individuals from starvation: given the low safety nets and salaries in many poor countries, lack of employment can severely reduce dietary intake resulting in other serious health-related issues or death. In that case, we recommend adding a death variable associated to forced unemployment. Such an approach exceeds the scope of this paper.

We present a simple model that incorporates the epidemiological model with the fiscal impact tied to mortality implications under different physical distance scenarios and we explore this numerically. For mathematical models that study in deep the mathematical models and specifically study stability and behavior of the non-linearity of the incidence and non-pharmaceutical intervention effect there are newly published papers of COVID-19 models that study these aspects see Bajiya et. al   \cite{bajiya2020mathematical} and  Bugalia et. al. \cite{bugalia2020mathematical}.  In these articles they study compartmental epidemic model incorporating quarantine and isolation to study patters of transmission as well as the efficacy of lockdown measures to lower transmissions rates.  In Bajiya, Bugalia, and Tripathi \cite{bajiya2020mathematical} they specifically show that the implementation of flawless isolation and almost 30\% of contact tracing the cumulative confirmed cases could be lowered to more than 50\% this result is important for the implementation of public health strategies for this or future pandemics as well as the proposed strategies we are presenting in our modeling approach. To study more on autonomous and non-autonomous epidemic models with nonlinear incidence rate see the work of Tripathi and Abbas \cite{tripathi2016global} where they study the stability of the disease free and the endemic equilibrium of this type of models and study the permanence, existence, uniqueness and asymptotic stability of an almost periodic solution of the model.  These kinds of studies are beyond the scoop of our paper that attain to study numerically different scenarios in the case of COVID-19, to minimize human and fiscal costs.

\section{Acknowledgments}

Authors are very grateful to three anonymous reviewers, the editor, Dr. Ricardo Gonz\'alez-M\'endez from the University of Puerto Rico School of Medicine as well as Dr. Ricardo J. Cordero-Soto from California Baptist University for their insights and suggestions. The usual disclaimers apply. 

\section{Disclosure}
\begin{itemize}
\item Both authors contributed equally in the design, codification and preparation of the article. Both authors have approved the final article.

\item This research did not receive any specific grant from funding agencies in the public, commercial, or not-for-profit sectors.

\item Declarations of interest: none
\end{itemize}

\bibliographystyle{apalike}
\bibliography{references}

\appendix
\section{Physical Distance Mechanism}

Let's look at Figure \ref{22six} to explain the mechanism of the physical distance modulation.  In this graph we vary the time length and the ratio of the infection to simulate the effect of alternating physical distancing measures. The top panel on Figure \ref{22six} is the dynamic of the symptomatic cases {\textbf{I}}, where the black plot represents the scenario where no measures are applied. Hence, there is no physical distancing $f(t) = 1$ for all $t$ to account for $100$\% contact rate. The gray graph is the one applying physical distancing as shown in the bottom graph that represents our piece-wise function $f(t)$. 

\begin{figure}[h]
\centering 
\includegraphics[width=\textwidth]{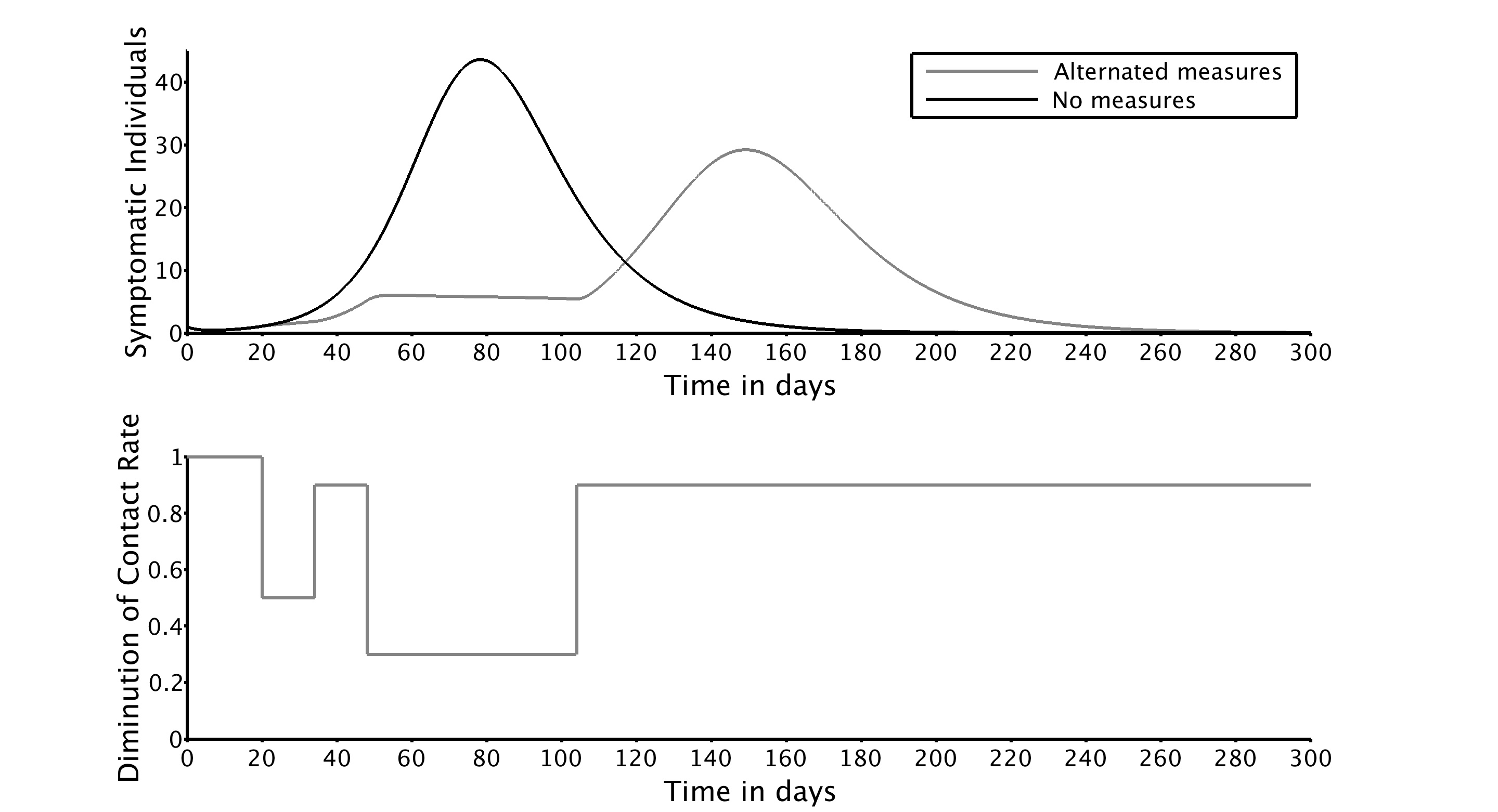}
\caption{The effect of alternate physical distancing measures on the epidemic curve.}
\label{22six}
\end{figure}

In this example we start physical distancing measures on day 20 (i.e. $t_{pdOn}=20$ after the start of the epidemic) and lower the transmission rate 50\% for two weeks (i.e. $t_{pdOff}=34$ two weeks after implemented + $t_{pdOn}=20$).  Then, measures are relaxed for two weeks. At this time the infection rate is decreased by 10\%, assuming people are a little bit more careful by washing their hands or using cloth masks.  After these two weeks, when cases start to increase again, the physical distancing measures are taken more strictly, thereby decreasing the infection rate by 70\% for six weeks.  Thus, we found the famous "flatten the curve" scenario in which we all desire to maintain our health system unsaturated. If a vaccine or effective treatment is not implemented before the physical measures are lifted, the epidemic will raise again.}

\section{Other Physical Distance Scenarios}

\begin{figure}[h]
\centering 
\includegraphics[width=\textwidth]{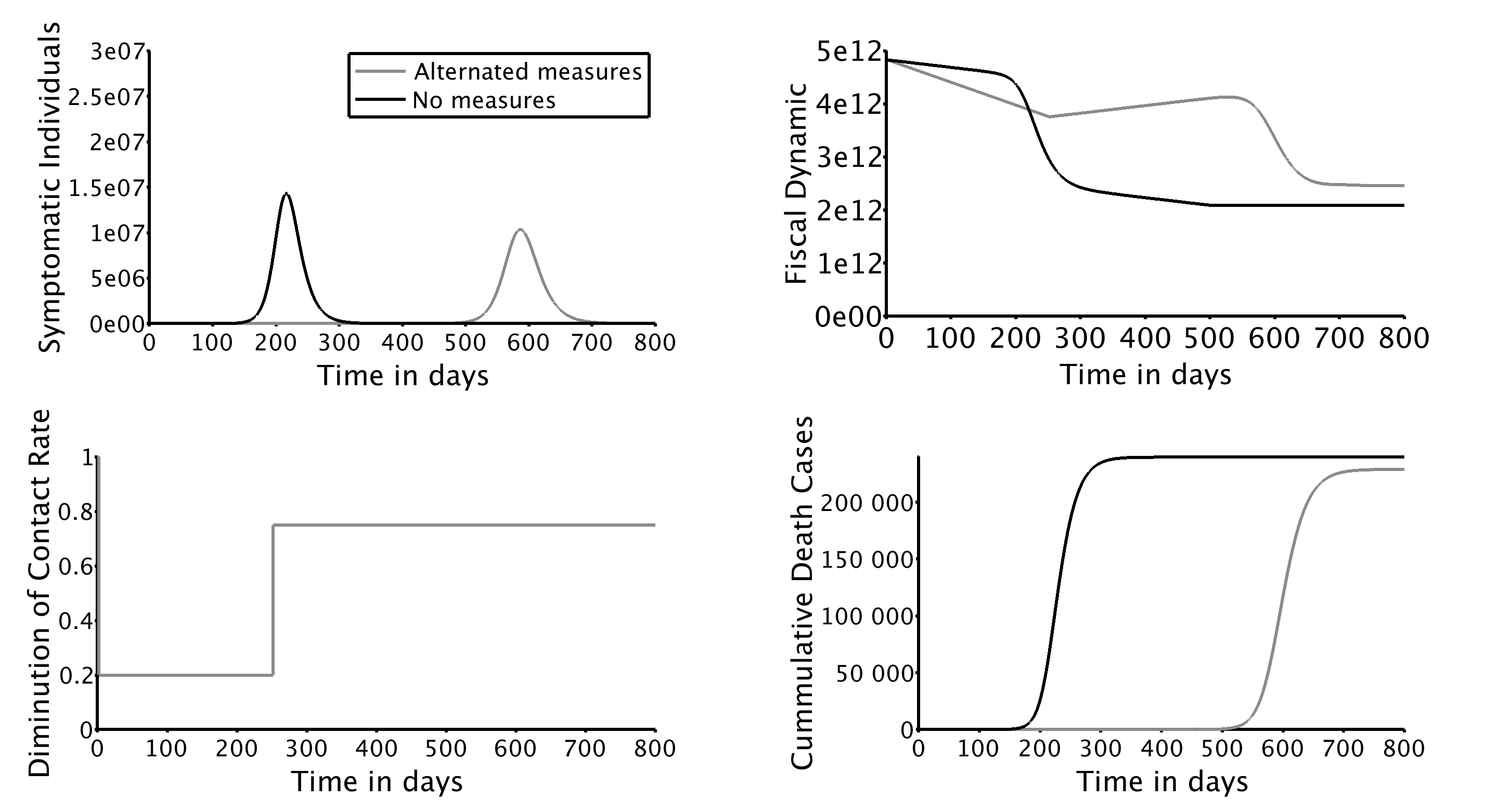} 
\caption{Distancing of 20\% for 250 days after the pandemic, followed by 75\% distancing.}
 \label{Fig:Ebola35}
\end{figure}

A policymaker would like to find an easier mechanism than a cyclical regime to combat the pandemic. In Figure \ref{Fig:Ebola35} we show the case where just one restrictive policy that declines the infection rate by 80\% is applied for 250 days. The probability of infection is then kept by a permanent reduction of 25\% to account for people being more careful and taking measures to avoid being infected. In this case the peak of the distancing curve arrives much later than before at approximately day 600. As expected, more lives (26 persons) are saved than in the alternated cycles of Figure \ref{Figure:EbolaCase_LD_02_day2_50First_75Others_4On_8off} but now the federal budget is almost halved. Note that in any of these scenarios, no physical distance or herd immunity is an optimal solution, either in terms of human or fiscal cost.

\begin{table}[h]
\begin{center}
\begin{tabular}{|c|r|r|}
\hline
\begin{tabular}[c]{@{}l@{}}Reduction in probability of infection\end{tabular} & \begin{tabular}[c]{@{}l@{}}Cumulative \\ Death cases\end{tabular} & Budget after 800 days \\ \hline
No measures    & 239,646 	 & \$2,090,225,896,159   \\ \hline
20\% for 250 days then 75\%                                    
& 228,930 	 & \$2,460,890,014,613   \\ \hline
\end{tabular}
\caption{Cumulative death cases and fiscal impact of the regime in Figure \ref{Fig:Ebola35}}
\label{Table:Ebola35}
\end{center}
\end{table}

\end{document}